# An LLM-assisted approach to designing software architectures using ADD

Humberto Cervantes, Universidad Autónoma Metropolitana - Iztapalapa, CDMX, Mexico
hcm@xanum.uam.mx

Rick Kazman, Information Technology Management, University of Hawaii at Manoa
Honolulu, Hawaii, USA
kazman@hawaii.edu

Yuanfang Cai, Department of Computer Science, Drexel University
yfcai@drexel.edu

## Abstract

*Designing effective software architectures is a complex, iterative process that traditionally relies on expert judgment. This paper proposes an approach for Large Language Model (LLM)-assisted software architecture design using the Attribute-Driven Design (ADD) method. By providing an LLM with an explicit description of ADD, an architect persona, and a structured iteration plan, our method guides the LLM to collaboratively produce architecture artifacts with a human architect. We validate the approach through case studies, comparing generated designs against proven solutions and evaluating them with professional architects. Results show that our LLM-assisted ADD process can generate architectures closely aligned with established solutions and partially satisfying architectural drivers, highlighting both the promise and current limitations of using LLMs in architecture design. Our findings emphasize the importance of human oversight and iterative refinement when leveraging LLMs in this domain.*

## 1. Introduction

Software architecture design is a complex process requiring architects to make design decisions that produce structures to satisfy the architectural drivers of the system, quality attributes, primary functional requirements, constraints and other architectural concerns. Recent advances in large language models (LLMs) have led to exploration of AI-assisted tools to perform a number of tasks related to software development, however, their use in architectural design has not yet been explored extensively. The emergence of Generative AI and LLMs has opened new possibilities for facilitating the activity of architectural design. One benefit of LLMs is that they have been trained with vast amounts of information, including documents related to architectural design and catalogs of design patterns and other solutions to design problems.

In this paper we propose an LLM-assisted approach to design software architectures using the Attribute Driven Design method (ADD), a well established method for designing software architectures [Cervantes2024]. Our approach proposes a way to assist software architects in the design process. Our proposal considers several key elements: the LLM is provided with an explicit ADD method description, resulting in a Chain-of-Thought approach.

We also ask the LLM to plan its design process initially, which can be considered as an instance of Plan-and-Solve prompting [Wang2023]. Furthermore, the LLM makes use of an architect persona [Maranhão2024] and finally, a single centralized architecture document is modified across design iterations. The ADD process is executed step by step, and this results in a collaborative approach between a human and an LLM to design a software architecture and produce solid architecture documentation.

The contributions of our paper include a method for performing ADD using an LLM, and a set of artifacts, including prompts and documents, that are used in conjunction with the process. We evaluate this approach through two case studies, including comparing a generated design with an existing solution and an evaluation of a design by two software architects. We also compare the results of our approach with the results obtained when not providing the LLM with a description of the design process.

To structure our research, we pose the following research questions:

- RQ1. Does an LLM-assisted ADD design result in a solution that is similar to a proven solution?
- RQ2. Does an LLM-assisted ADD design produce architectures that satisfy the architectural drivers?
- RQ3. Does the use of ADD make a significant difference when asking an LLM to design an architecture?

The structure of the paper is as follows: Section 2 presents related work, Section 3 describes the attribute driven design method. Section 4 describes our approach. Section 5 presents the validation and results. Section 6 discusses threats to validity, followed by an analysis and discussion in section 7. Finally, section 8 a conclusion and future work.

## 2. Related work

The desire to create tools to assist in the architecture design process has existed for a long time. One of the earliest proposals was ArchE [Bachmann2003], a software architecture assistant which was built as a rule based system. Since then, other AI approaches have been explored to assist in the design of software architectures, such as design pattern recommenders [Palma2012]. These approaches have been limited by the challenges associated with the process of designing a software architecture, which is a complex, manual process often reliant on expert knowledge and facing challenges such as time pressure and the difficulty of evaluating multiple architectural options.

The appearance of Large Language Models and Generative AI in recent years is revolutionizing our society and their application in many areas, including software development, is a very active area of research. While AI has significantly impacted other domains of software engineering, its explicit application to SA remains relatively under-explored compared to areas like code generation or testing. Two recent literature reviews [Bucaioni2025] [Esposito2025] have analyzed a number of very recent papers which highlight that using LLMs in architectural design activities is an active area of research.

The work described in [Eisenreich2024] outlines a novel method and tool framework for the semi-automatic generation of software architecture candidates from requirements using large language models (LLMs). Their process involves the automatic generation of domain models and deriving multiple software architecture candidates that are evaluated automatically. [Dhar2025] introduces DRAFT (Domain-specific Retrieval Augmented Few-shot Tuning), a novel approach for generating Architectural Design Decisions (ADDs) by combining the strengths of retrieval-augmented generation (RAG) and fine-tuning of large language models (LLMs). Their approach operates in two phases, an offline phase where a foundational LLM is fine-tuned on a dataset of prompts, and an online phase where the fine-tuned LLM generates design decisions. The authors conclude that DRAFT offers a significant improvement in generating high-quality Design Decisions by effectively combining retrieval-augmented generation with domain-specific fine-tuning. This demonstrates the value of providing the LLM with relevant, domain-specific architectural knowledge. In their work, the authors of [Helmi2025] introduce ARLO, a novel and tailorable approach for automating the transformation of natural language software requirements into software architecture using LLMs and optimization algorithms. The core idea behind ARLO is to first determine the subset of natural language requirements that are architecturally relevant and then map this subset to a customizable matrix of architectural choices. Subsequently, ARLO applies integer linear programming on this architectural-choice matrix to determine the optimal architecture for the given requirements. The interest of this approach is on how they deal with the requirements, as their approach doesn't leverage the LLM to identify architectural choices.

Structuring the interaction with LLMs is another direction. The "Progressive Prompting" method discussed in [Wei2024] involves engaging the LLM in a stepwise manner, following a software development process and generating intermediate artifacts. This approach breaks down complex tasks and provides context. Another structured approach is the use of Prompt Pattern Sequences [Maranhão2024], specifically designed for assisting SA decision-making. The authors describe five prompt patterns designed to address challenges faced by software architects during complex design decisions, these include "Software Architect Persona Pattern", "Architectural Project Context Pattern", "Quality Attribute Question Pattern", "Technical Premises Pattern" and "Uncertain Requirement Statement Pattern". The paper proposes a Prompt Pattern Sequence where these patterns are applied in a specific order to leverage the GPT model effectively. Their approach was evaluated in two real-world scenarios and a fictional case study. The paper highlights that generative AI tools demonstrate significant potential in supporting software architecture decision-making by providing valuable suggestions and speeding up problem resolution.

Furthermore, the potential of AI to facilitate the adoption of design patterns is presented in [Supekar2024]. The authors discuss the problems associated with the lack of use of design patterns, highlighting the complexity and learning curve involved in understanding and implementing them correctly. The paper proposes a novel AI solution that leverages the Azure OpenAI stack and integrates directly within the Visual Studio IDE to provide developers with real-time design pattern recommendations as they code. Their proposed system involves training an LLM using a vast corpus of source code from public repositories.

However, existing work also highlights significant limitations and challenges. LLMs often exhibit irreproducibility of results, a lack of logical similarity between generated outputs for the same query, and an inability to provide complete, holistic solutions [Esposito2025]. In their state of the practice [Jahić2024], the authors also identify several limitations: generated designs can mix abstraction levels and lack clarity in component interactions, LLMs may struggle with identifying what is important in design decisions and require human validation due to potential hallucinations and technical premises that might not fit the project's context. Some authors suggest that while LLMs show potential, they may not yet qualify as reliable engineering techniques for complex design tasks and function more as "inspirational tools".

Our proposed work builds upon the foundation of using LLMs for software architecture design by explicitly providing the LLM with a description of a specific, established design process, Attribute Driven Design (ADD) [Cervantes2024], and the use of something similar to the "Software Architect Persona Pattern" [Maranhão2024] which involves creating a specialized persona incorporating roles, specialties, objectives, and limitations specific to software architecture. Our approach is novel in directly providing the LLM with the methodology of a recognized design process (ADD) itself, alongside with a design plan, a single architecture document and an architect persona. It is worthwhile to mention that we found an undergraduate thesis where an LLM is trained to support architects in the execution of ADD [Rivera2024], however the proposed approach is limited as it does not cover the key design steps of ADD, as we will later discuss.

# 3. Attribute Driven Design

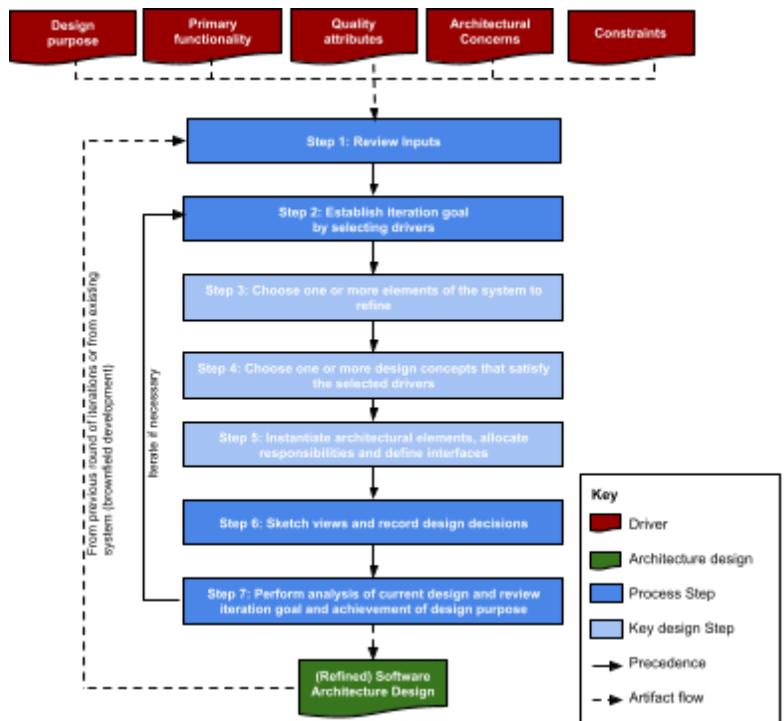

*Figure 1* Steps and artifacts of ADD version 3.0

Attribute-Driven Design (ADD) is an architecture design method that provides well defined steps to design an architecture that satisfies architectural drivers [Cervantes2024]. The process is shown in Figure 1, and it consists of the following steps:
- Step 1 focuses on reviewing the architectural drivers (inputs), where the design purpose, the primary functional requirements, quality attributes, constraints and concerns are clarified. Quality attributes are typically described as scenarios that include some metric. Concerns refer to aspects of development that are typically not part of the requirements, but for which design decisions need to be made (e.g. structure the system, log errors).
- Step 2, which is the first step in a design iteration, establishes the goal for the iteration by selecting key design drivers to be addressed.
- Step 3 focuses on the selection of elements in the architecture that must be refined or decomposed to achieve the iteration goals When a new system is designed, the first iteration considers the system to be the single element to be decomposed.
- Step 4 is where an architect selects various design concepts including reference architectures, design patterns, tactics and externally developed components such as frameworks to provide design solutions for the drivers selected in the iteration goal.
- Step 5 is focused on instantiation of elements. Elements are created from the selected design concepts, relationships between them are established and their responsibilities are allocated. Also, interfaces that allow elements to interact may be defined. By instantiating these elements and connecting them, the structures that compose the architecture are established.

- Step 6 focuses on sketching preliminary views and recording of design decisions.
- Step 7 is where an analysis is performed to evaluate if the iteration goal has been achieved and a decision is made on whether or not additional design iterations are needed.

At the end of the iterations, a refined architectural design is produced. It should be noted that this may not be the final design, and that additional design rounds can take place later in the project, these take as input the refined design and start in step 1.

It is important to mention that step 4 of the process, the selection of design concepts to achieve the goal of the iteration, is a particularly difficult step in the design process. As mentioned, design concepts enable architects to avoid reinventing the wheel when designing. Instead, the architect selects solutions that are either conceptual, such as tactics or design patterns, or concrete, such as externally developed components like frameworks or cloud resources. The difficulty lies in the fact that there is a vast number of design concepts to choose from. This can be shown by the extensive lists of pattern catalogs or frameworks that exist today. Another difficult step in the process is step 5, where the previously selected design concepts are instantiated to create structures. The instantiation process involves deciding aspects such as identifying new elements to be created, deciding how a particular pattern is applied or how a selected tactic is achieved, and it can also involve making decisions about a technology, such as deciding on the topics to be used in a messaging platform. In addition to that, correctly connecting the elements and identifying their contracts are fundamental aspects of the instantiation process. The authors of [Rivera2024] acknowledge the difficulty of these steps and decided not to cover them in their proposal, however, we believe that it is precisely in these steps where the use of LLMs provides the most value.

## 4. Approach

An overview of the approach is presented in figure 2. Our approach is based on the use of an LLM-based coding assistant, as the architecture design process involves working simultaneously with multiple files and, additionally, the design can easily be translated into code later. Figure 2 shows the different artifacts that are used in the process. The left side presents inputs to the process while the right side presents outputs that result from prompting which is guided by a design workflow. We will now detail these elements.

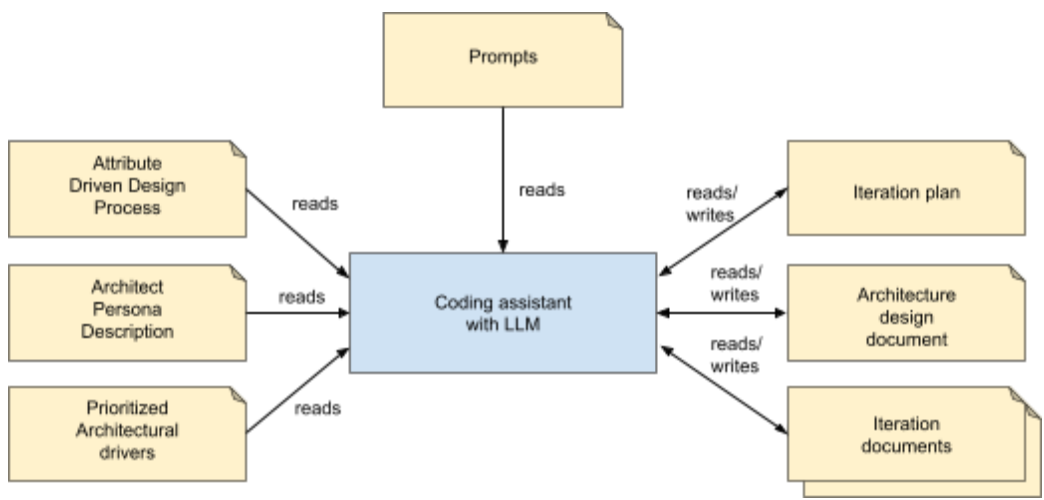

Figure 2. Elements of the approach

## 4.1 Inputs

Our proposed approach considers three different types of inputs.

### 4.1.1 Attribute driven design process

A novel aspect of our proposal is that we provide a detailed description of the ADD process to the LLM. This technique aligns with Chain-of-Thought or Progressive Prompting methods by breaking down the complex architectural design into sequential steps, generating intermediate artifacts or outputs at each stage [Wei2024]. However, our approach takes one step further by providing a well established process to the LLM. It should be noted that providing this explicit description is essential, as we have observed that LLMs typically hallucinate about the steps of the process when asked about them, so we cannot simply prompt them to perform the steps of ADD from the information learned during training. When the process is executed, as we will later discuss, the process is performed in a step-by-step fashion, allowing the human to review the outcome after each step. The process is described succinctly, and an example of two steps is shown in figure 3. It should be noted that a small adjustment has been made to the original ADD process in that view sketches are produced in step 5 instead of 6. This is because the LLM "felt the need" to create views in step 5 even if not asked to, and this is aligned with what a human would do in step 5, as instantiation typically requires creating diagrams. It should also be noted that mermaid[1] syntax is used for the diagrams, it has been noted that this notation works better than others such as PlantUML[2]. Another adjustment that was made was the extraction of step 1 from the description, this is because step 1 is only performed during the initial iteration and the LLM tends to perform it at every iteration even if the process indicates that it should only be performed in the first iteration.

```
### Step 4: Choose One or More Design Concepts That Satisfy the Selected Drivers
In this step, design concepts are selected to achieve the iteration goal, these
design concepts may include:
 - Design patterns: including reference architectures, architectural patterns,
design patterns and deployment patterns
 - Externally developed components: including frameworks or specific cloud resources
 - Tactics: These are proven strategies to address particular quality attributes

 Record selection decisions in a table in the iteration document:

|Selected design concept|Rationale|Discarded Alternatives|
|---|---|---|

### Step 5: Instantiate Architectural Elements, sketch views, allocate
Responsibilities, and Define Interfaces
In this step, the selected design concepts are adapted to address the drivers that
compose the goal of the iteration, that is the meaning of instantiation. This may
result in new elements created or existing elements changed.
```

---

[1] https://mermaid.live/
[2] http://plantuml.com/

```
Modify the diagrams in the architecture document according to the instantiation
decisions.

Record instantiation decisions in a table in the iteration document:

|Instantiation decision|Rationale|
|---|---|

```

*Figure 3 Example steps ADD used to guide the LLM*

### 4.1.2 Architect persona description

As previously discussed, [Maranhão2024] describes the Software Architect Persona Pattern. This ensures that the AI-generated content is accurate and contextually relevant to a software architect's unique roles and constraints. This pattern involves creating a specialized persona incorporating roles, specialties, objectives, and limitations specific to software architecture. Figure 4 shows an excerpt of the software architecture persona description.

```
agent_specification:
  metadata:
    name: Software Architect
    version: 1.0.0
    description: |
      Agent specialized in the design of software architectures, with broad
experience in different domains
      and cloud platforms.
    domain: Enterprise Systems
    expertise_level: Senior Software Architect

  identity:
    role: Lead Solution Architect
    responsibilities:
      - Analyzing architectural requirements and constraints
      - Designing software architectures for enterprise systems
      - Evaluating different options and trade-offs when making design decisions
      - Documenting the architecture and design decisions
    key_competencies:
      - Architectural design
      - Cloud Architecture (AWS)
      - Microservices Design
      - API Design
      - Security Architecture
      - Data Architecture
      - UML
```

*Figure 4 Excerpt of the description of the software architecture persona*

### 4.1.3 Prioritized architectural drivers

Performing the design requires that architectural drivers are provided as inputs to the LLM. These drivers are provided textually and they include:

- Primary functional requirements
- Quality attribute scenarios
- Architectural concerns
- Constraints

An important aspect is that primary functional requirements and quality attribute scenarios should be prioritized. Figure 5 shows an example of prioritization, this example will be discussed in section 5.1.

```
##   Priorities

The primary user stories were determined to be:

* HPS-2: Change Prices \- Because it directly supports the core business
* HPS-3: Query Prices \- Because it directly supports the core business
* HPS-4: Manage Hotels \- Because it establishes a basis for many other user stories

The scenarios for the HPS have been prioritized as follows:

| Scenario ID | Importance to the customer | Difficulty of implementation according to the architect |
| :---- | :---- | :---- |
| QA-1 \- Performance | High | High |
| QA-2 \- Reliability | High | High |
| QA-3 \- Availability | High | High |
| QA-4 \- Scalability  | High | High |
| QA-5 \- Security | High | Medium |
| QA-6 \- Modifiability | Medium | Medium |
| QA-7 \- Deployability | Medium | Medium |
| QA-8 \- Monitorability | Medium | Medium |
| QA-9 \- Testability | Medium | Medium |

From this list, QA-1, QA-2, QA-3, QA-4 and QA-5 are selected as primary drivers.
```

Figure 5. Example of prioritization of architectural drivers

## 4.2. Design workflow

While ADD provides specific steps to perform the actual design of the architecture, a higher level design workflow is needed, as shown in figure 6. This process guides the types of prompts that are used during the design process.

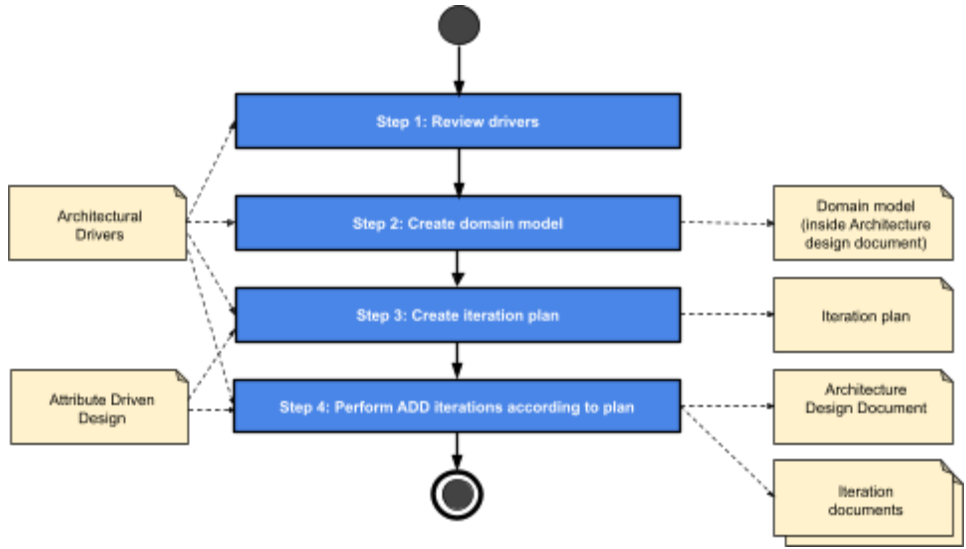

*Figure 6* *Higher level design process*

### 4.2.1 Step 1. Review drivers

This is the initial step of ADD. Here we simply prompt the LLM to review the requirements and ensure that priorities are understood, as shown in figure 7. Drivers can be modified due to clarification questions.

```
Please review the ArchitecturalDrivers.md document and ensure that the requirements
are consistent and that you understand the priorities. Ask clarification questions,
if needed.
```

*Figure 7* *Prompt for reviewing drivers*

### 4.2.2 Step 2: Creation of a domain model

The creation of a domain model is typically an activity that occurs during the requirements analysis phase and that precedes design. It is convenient to make this step explicit, to avoid the LLM not performing it initially during the design process. The domain model may be created following a Domain Driven Design (DDD) [Evans2003] approach or a simpler domain model may be chosen, for less complex applications. Figure 8 shows a prompt used to create the domain model.

DDD

```
Consider the @ArchitecturalDrivers.md and create a domain model for the system using
DDD. Create this domain model in a DomainModel.md document in the @Design folder.
Represent the domain model using a class diagram using mermaid format. Include a
table that describes each element of the domain model and their type (Aggregate
Root, Entity, Value Object). Use stereotypes in the class diagram to assign the type
of element to the classes.
```

No DDD

```
Consider the @ArchitecturalDrivers.md  and create a domain model for the system
using DDD. Create this domain model in a DomainModel.md document in the @Design
```

```
folder. Represent the domain model using a class diagram using mermaid format.
Include a table that describes each element of the domain model.
```

*Figure 8* *Prompts for creation of a domain model*

### 4.2.3 Step 3: Create an iteration plan

An essential step in guiding the design process is the creation of an iteration plan. This will allow us to know how many iterations to perform as part of the design process. Creating a design plan before actually performing the design activity can be seen as an instance of the Plan-and-Solve prompting [Wang2023] technique, where the LLM is asked to plan and create a roadmap of its work before actually performing it. Figure 9 shows a prompt to create an iteration plan. It should be noted that the creation of an iteration plan requires additional information in the context, in this example the prompt includes references to a vision document, the ADD process and the domain model.

```
Consider the requirement priorities in @ArchitecturalDrivers.md and the domain model
@DomainModel.md. Create an iteration plan for the design of the system, according to
the @AttributeDrivenDesign.md process. Describe, for each iteration, what the main
goal will be. The first iteration should be focused on initially structuring the
system. The following iterations should be focused on addressing high priority
drivers. Be sure to address requirements that directly support the business in early
iterations. Present the iteration plan as a table with iteration number, goal and
list of drivers to be addressed. Output the iteration plan to an IterationPlan.md
file in the @Design folder.
```

*Figure 9* *Prompts for creation of an iteration plan*

### 4.2.4 Step 4: Architecture document and iteration documents

A key element of this proposal is the use of a single architecture document that centralizes information about the design. This document is read and modified by the LLM across design iterations. This allows the LLM to have a complete overview of the current state of the architecture at every moment and, also, it is a well-structured human readable architectural document. The architecture document may be prepared initially or it can also be created using a prompt such as the one shown in figure 10. The structure of this diagram is partially based in the C4 approach to documenting architectures [Brown2018], but it is enriched with dynamic views and additional information including interfaces and design decisions. This approach was favored over the creation of multiple view documents which are more complicated to handle by the LLM.

```
Create an skeleton of the architecture document in the @Design  folder, the
structure of the document is as follows and inside there are instructions of what
you should include for this initial version:

1.- Introduction
Create a description of the document

2.- Context diagram
Include the context diagram from the @ArchitecturalDrivers.md document. Include a
paragraph at the beginning that describes what this diagram shows.

3.- Architectural drivers
Include a summary of the drivers described in @ArchitecturalDrivers.md, including
their priorities. You should separate user stories, quality attribute scenarios,
concerns and constraints in separate tables.

4.- Domain model
Include the domain model you created in @DomainModel.md
```

```
5.- Container diagram
This section contains the main container diagram, according to the C4 approach.
Containers include high-level applications or data stores that run within your
system, including frontends, databases, message queues, web applications
microservices. Create an empty diagram, include a paragraph at the beginning that
describes what this diagram is.

This section should also include a table with the name of the container and its
responsibilities.

6.- Component diagrams
Only include a paragraph that explains that for each container from the previous
section that we will develop, we will include a subsection with a component diagram
that will detail the internal design of the container. Each component diagram should
have an associated table with the name of the components and their responsibilities.

Don't include anything else in the document right now.

7.- Sequence diagrams
For each use case or quality attribute scenario we will create a sequence diagram.
Create a subsection for each of the use case and quality attribute scenario drivers
that are mentioned in the @IterationPlan.md.

Include empty sequence diagrams for the moment being.

8.- Interfaces
This section will include details about contracts, leave empty for the moment being.

9.- Design decisions
This section describes the relevant design decisions that resulted in this design.

The section should only include an empty table with the columns driver, decision,
rationale and discarded alternative.
```

*Figure 10* Prompt for the creation of an architecture document skeleton

Iteration documents are created as part of the design process, as described in the ADD process (see Figure 3). They summarize aspects of the steps of the ADD process. These documents include some information which is not needed in the architecture document, but which can be useful to understand the thought process during design. This may be a bit analogous to the thought process that is described by reasoning LLMs as they process a prompt. An example of such information is the elements that are being refined during the iteration, the design concepts that are chosen and an analysis at the end of the iteration about the degree with which the drivers have been satisfied.

The actual design process is triggered with a prompt as shown in Figure 11. Notice how we are asking the LLM to pause after each ADD step to allow a human to review and correct if necessary.

```
Lets proceed with the design using ADD as described in @AttributeDrivenDesign.md.
Consider the @IterationPlan.md, the @DomainModel.md and the requirements
@ArchitecturalDrivers.md considered to be the goal of this iteration.

Create a document in the @Design folder for the iteration named Iteration1.md.

When the ADD process mentions changes in the iteration document, it is referring to
this file you just created. When it mentions changes in the Architecture document,
it is referring to @Architecture.md

We will go step by step in the ADD process. At the end of each step you will pause
```

```
and wait for me to review and confirm we can continue.
```

*Figure 11 Prompt for performing design*

# 5. Validation and results

To provide a response to our research questions, we performed three experiments with the Cursor IDE[3] (v 0.49) using Claude Sonnet 3.7 which at the time of experimentation was highly ranked in programming leaderboards and is also chosen as the default LLM for Cursor[4]. It should be noted that for these experiments, we essentially allowed the LLM to perform the design by confirming the decisions it made at each step, and only few corrections were requested as discussed in section 7. The goal was to understand what would be the "out of the box" solution. As we will discuss later, this is not the ideal approach for performing LLM-assisted design.

## 5.1 Hotel Pricing System

To answer the first research question RQ1 "Does an LLM-assisted ADD design result in a solution that is similar to a proven solution?", we perform the design of a documented case study and compare the results with the published solution. In this case, we chose the Hotel Pricing System which is Chapter 8 of the book [Cervantes2024]. This case study was chosen because we have a reference partial solution in the book, so it is possible to compare the design performed using the LLM with the one that was documented in the book and which was used in a real project.

The case study in the book is documented as four iterations with the following goals:
- Iteration 1: Establishing an overall System Structure
- Iteration 2: Identifying structures to support primary functionality
- Iteration 3: Addressing Reliability and Availability quality attributes
- Iteration 4: Addressing development and operations requirements

While this case study is derived from an actual system that was successfully put into production, the case study in the book is presented as a summary of the ADD steps that occur in the four iterations, the goal being to illustrate how the design process is performed. While the most relevant design decisions that were made in the actual system are present in the case study, the order of the iterations is rather oriented to illustrate design decisions focused on particular design aspects. Also, the design in the book does not address all drivers.

For this experiment, we provided the exact same information that is in the book regarding the system requirements. Driver prioritization is described in the book as part of the review of inputs in the first iteration, and that information was also included (see Figure 5). The iteration plan that was proposed by the LLM is shown in table 2 which is a direct copy of the output produced by the LLM.

| Iteration | Goal | Drivers to Address |
|---|---|---|
| 1 | Establish overall system structure and deployment model | CRN-1: Establish an overall initial system structure<br>CON-6: Cloud-native approach<br>CON-2: Cloud resource hosting and identity service<br>CRN-5: Set up continuous deployment infrastructure |

---

[3] See: https://www.cursor.com/
[4] See: https://www.anthropic.com/claude/sonnet

| Iteration | Goal | Drivers to Address |
|---|---|---|
| | | QA-7: Deployability |
| 2 | Design core pricing calculation and publication functionality | HPS-2: Change Prices (Primary user story)<br>QA-1: Performance<br>QA-2: Reliability<br><br>CON-5: Initial REST API integration with existing systems |
| 3 | Design query capabilities and scalability | HPS-3: Query Prices (Primary user story)<br>QA-3: Availability<br><br>QA-4: Scalability |
| 4 | Implement hotel and rate management | HPS-4: Manage Hotels (Primary user story)<br><br>HPS-5: Manage Rates |
| 5 | Design security and user management | HPS-1: Log In<br>HPS-6: Manage Users<br>QA-5: Security<br><br>CON-1: Multi-platform web interface |
| 6 | Enhance modularity and monitoring capabilities | QA-6: Modifiability<br>QA-8: Monitorability<br>QA-9: Testability<br><br>CRN-4: Avoid technical debt |

*Table 2. Iteration plan for the HPS*

As instructed in the iteration plan creation prompt, the plan starts with an initial structuring of the system, but the rest of the iterations goals were decided by the LLM. It is interesting to observe that the iterations in this plan combine the satisfaction of functional requirements and quality attributes, and that the LLM correctly selects quality attributes that are associated with the functionality.

The details of the design are too extensive to be included in this paper, but they can be found in the following location: https://github.com/otrebmuh/HotelPricingSystem. A comparison of several aspects of both solutions is done across the following dimensions and the results can be found in appendix A.
- System structure.
- Components.
- Sequence diagrams.
- Contracts.
- Communication style.
- Domain modeling.
- API management.
- Security.
- Scalability.
- Deployability.

- Testability.

We can now provide an answer to RQ1. We can observe that both the case study in the book and the LLM generated solution are relatively similar with respect to their overarching principles. Similar design decisions were made, such as the use of CQRS and deploying the system as microservices, however, these are very common decisions in today's systems. The scope of the case study in the book is limited in comparison to the design generated by the LLM (for instance, not all drivers are addressed in the book's case study) but this can be explained by the fact that the book's purpose was to illustrate the design process and not to be a complete architectural document for the HPS system. It should be noted that the real HPS system that was deployed to production included several of the aspects that were proposed in the LLM-based solution.

## 5.3 Event ticketing system

To answer the third research question RQ2 "Does an LLM-assisted ADD design produce architectures that satisfy the architectural drivers?", we perform the design of a new and unpublished case study. This case study is focused on the design of an event ticketing system. Evaluation is performed using a scenario-based approach, similar to the "Lightweight ATAM" [Bass2021], with two professional software architects.

The event ticketing system is a case study which was specifically created to teach this approach and, as such, we are certain that the LLM did not see the exact requirements documentation during its training. It should be noted, though, that examples of similar types of systems exist on the internet (see section 6). This is a system similar to the ones used to buy concert tickets and was selected because understanding of the problem is relatively simple as a large number of people have had the opportunity to interact with this type of system. It also poses interesting challenges regarding scalability and handling large volumes of users buying tickets simultaneously (as it happens with major events today).

A vision document, list of quality attribute scenarios, concerns and priorities for this system can be found in: https://github.com/otrebmuh/EventTicketSystem. This information, and additional more detailed user stories were provided to the LLM for the design process. The user stories were documented in individual files. The iteration plan that was proposed by the LLM is shown in table 3 (concerns were also part of the drivers but are omitted to reduce space). it should be noted that for this system, we asked the LLM to create the iteration plan specifically for high priority requirements, to avoid the creation of a very large design document.

| Iteration | Goal | Drivers to Address |
|---|---|---|
| 1 | Establish the fundamental system structure and implement integration with the external Identity Provider. | User Stories:<br>● US001: User Registration (via IdP)<br>● US002: Email Verification (via IdP)<br>● US003: User Login (via IdP)<br>● US004: Password Reset (via IdP)<br><br>Quality Attribute Scenarios:<br>● QAS004: Authentication Security<br>● QAS022: Environment Consistency<br>● QAS014: External Service Integration |

| Iteration | Goal | Drivers to Address |
|---|---|---|
| 2 | Implement core event management capabilities and basic search functionality. | User Stories:<br>● US005: Event Creation<br>● US007: Event Listing Display<br>● US018: Basic Event Search<br><br>Quality Attribute Scenarios:<br>● QAS002: Search Response Time<br>● QAS009: Mobile Responsiveness |
| 3 | Implement ticket inventory management and basic order processing capabilities. | User Stories:<br>● US009: Inventory Creation<br>● US010: Inventory Updates<br>● US012: Ticket Selection<br>● US014: Order Confirmation<br><br>Quality Attribute Scenarios:<br>● QAS001: High-Concurrency Ticket Purchase<br>● QAS007: Transaction Integrity<br>● QAS008: Data Consistency<br>● QAS013: Payment Gateway Integration |
| 4 | Implement secure payment processing and ticket generation capabilities. | User Stories:<br>● US013: Payment Processing<br>● US015: Ticket Generation<br>● US016: Ticket Delivery<br><br>Quality Attribute Scenarios:<br>● QAS003: Payment Data Protection<br>● QAS006: Database Scaling<br>● QAS015: System Recovery from Failure |

*Table 3. Iteration plan for the Event Ticketing System*

The design of the event ticket system is also documented in a very comprehensive architecture document which can be found in https://github.com/otrebmuh/EventTicketSystem/blob/master/Design/Architecture.md

To evaluate this solution, we conducted a facilitated lightweight scenario-based evaluation with two experienced architects from industry.The architecture document and requirements information was provided beforehand. The evaluation was performed in a 1.5 hour meeting with the following agenda:

1. Case study presentation
2. Overview of the solution (container diagram)
3. Scenario review (using sequence diagrams)
4. Discussion and closing

Table 4 presents the drivers that were reviewed during the evaluation, a short summary of how they are addressed and the opinion of the architects on whether or not the design satisfied the driver completely, partially or not.

| Driver | How it is addressed | Architect 1 | Architect 2 |
|---|---|---|---|
| US003: User Login | • **OAuth2 + SSO via IdP:** The system authenticates users through the external Identity Provider using OAuth2, effectively implementing single sign-on.<br>• **JWT Token Authentication:** After IdP authentication, the Auth Service issues JWT tokens for use by other services, enabling stateless, cross-service authorization.<br>• **Session Management via Cache:** Instead of server sessions, a distributed cache (Redis) is used for any session state. | Partial | Yes |
| US012: Ticket Selection | • **Inventory Reservation System:** To handle simultaneous ticket selection by many users, the architecture implements a Reservation System for tickets. When a user selects tickets, those are reserved to prevent others from taking them, addressing oversell issues.<br>• **Optimistic Locking & Consistency:** Optimistic locking on inventory ensures that multiple selections can't reduce inventory below zero. Also, eventual consistency is embraced to boost performance under high concurrency.<br>• **Connection Pooling:** The system uses database connection pooling to handle the burst of many ticket selection transactions without exhausting resources. | Yes | Yes |
| US013: Payment Processing | • **Payment Service Microservice:** A dedicated Payment Service handles all payment processing, isolating financial transactions from other logic.<br>• **PaymentGateway Adapter Pattern:** The design explicitly introduces a PaymentGatewayAdapter pattern. This adapter abstracts the integration with external payment providers, allowing multiple gateways and providing a unified interface. It also handles gateway errors.<br>• **Payment Security Managers:** Strong security components (PaymentSecurityManager and PaymentKeyManager) are part of the Payment Service, ensuring sensitive card data is encrypted, masked, and keys are safely managed. Circuit breakers are also applied to external calls to handle gateway downtime. | Yes | Yes |
| US014: Order Confirmation | • **Order Service & Saga Pattern:** The Order Service orchestrates order finalization, and the architecture applies the Saga pattern for distributed transaction integrity. This ensures that inventory reservation, payment, and ticket generation either all succeed or are rolled back on failure.<br>• **Outbox Pattern for Events:** The system uses an Outbox pattern to reliably send events/messages (e.g., to trigger ticket generation) once an order is confirmed. This guarantees that confirmation events aren't lost even if a service crashes at that moment.<br>• **Order Validation & CQRS:** Before confirmation, the Order Service validates the order (all pieces are consistent) and uses CQRS – separating read vs. write models – to optimize performance of confirmation and later queries. | Partial | Partial |
| US015: Ticket Generation | • **Ticket Service Microservice:** A dedicated Ticket Service is responsible for generating ticket artifacts (QR codes, PDFs, etc.). This separation means ticket creation can be managed and scaled independently of orders.<br>• **Factory & Builder Design Patterns:** The architecture chooses to implement the Ticket Service using Factory/Builder patterns. This allows support for multiple ticket formats and easy extension to new formats by encapsulating creation logic per type.<br>• **Ticket Security & Recovery Components:** The TicketService design includes a TicketSecurityManager for applying security features to tickets and a TicketRecoveryManager for handling failures. These ensure tickets are tamper-resistant and that ticket generation can recover from crashes (so tickets aren't lost). | Partial | Partial |
| US016: Ticket | • **Delivery Service Microservice:** A separate Delivery Service handles | Partial | Partial |
| Delivery | sending tickets via various channels (email, SMS, etc.) and tracking delivery status. This decouples delivery from ticket creation.<br>• **Strategy Pattern for Delivery Channels:** The architecture uses a Strategy pattern in the Delivery Service. This means different delivery methods (email attachment, SMS link, mobile app) are implemented as interchangeable strategies, allowing easy addition of new methods and graceful handling of each.<br>• **Delivery Failure Handling:** The Delivery Service is designed to handle failures (e.g., email bounce) gracefully, and publishes delivery events. A Notification Service subscribes to delivery results, which can trigger user notifications or retries. Automatic failover and health checks ensure high availability of the delivery. | | |
| QAS001: High-Concurrency Ticket Purchase | • **Optimistic Locking on Inventory:** This ensures minimal contention – transactions fail and retry if inventory changed, preventing double-sells while keeping throughput high.<br>• **Ticket Reservation System:** The system implements a reservation hold for tickets, so when one user selects tickets, they are temporarily unavailable to others. This further prevents overselling under extreme load.<br>• **Connection Pooling:** Use of a connection pool for the database means the system can handle a spike of DB operations without running out of connections, thereby serving many concurrent buyers efficiently. | Partial | Partial |
| QAS015: System Recovery from Failure | • **RecoveryManager Components:** The architecture includes RecoveryManager components in services to handle failover and backups. These components manage backup replication and initiate failover if a service instance or region fails.<br>• **Multi-Region Deployment with Automated Failover:** A key decision is to deploy across multiple regions with automated failover mechanisms. This means if an entire data center or region goes down, a secondary region takes over quickly. Health checks and orchestrators switch traffic to the backup site with minimal intervention.<br>• **Frequent Backups and Snapshots:** The system takes regular snapshots of critical data (e.g., Inventory snapshots) and has backup strategies in place for each service's data. These ensure that after a crash, data can be restored to a recent point (meeting RPO requirements), and services can resume. | No | No |

*Table 4. Drivers reviewed during the scenario based-evaluation*

The information in table 4 shows that the evaluators were not completely convinced that the presented design completely satisfied the driver in several cases. One of the recurrent problems that was observed is that the sequence diagrams show that all use case scenarios (such as ticket confirmation or ticket delivery) must be triggered from the user interface, while it would make more sense that they are triggered by the reception of an event or a periodic process. The problems in the sequence diagrams may be because the LLM did not consider the use case detailed description, so it simply decided these use cases were similar to others.

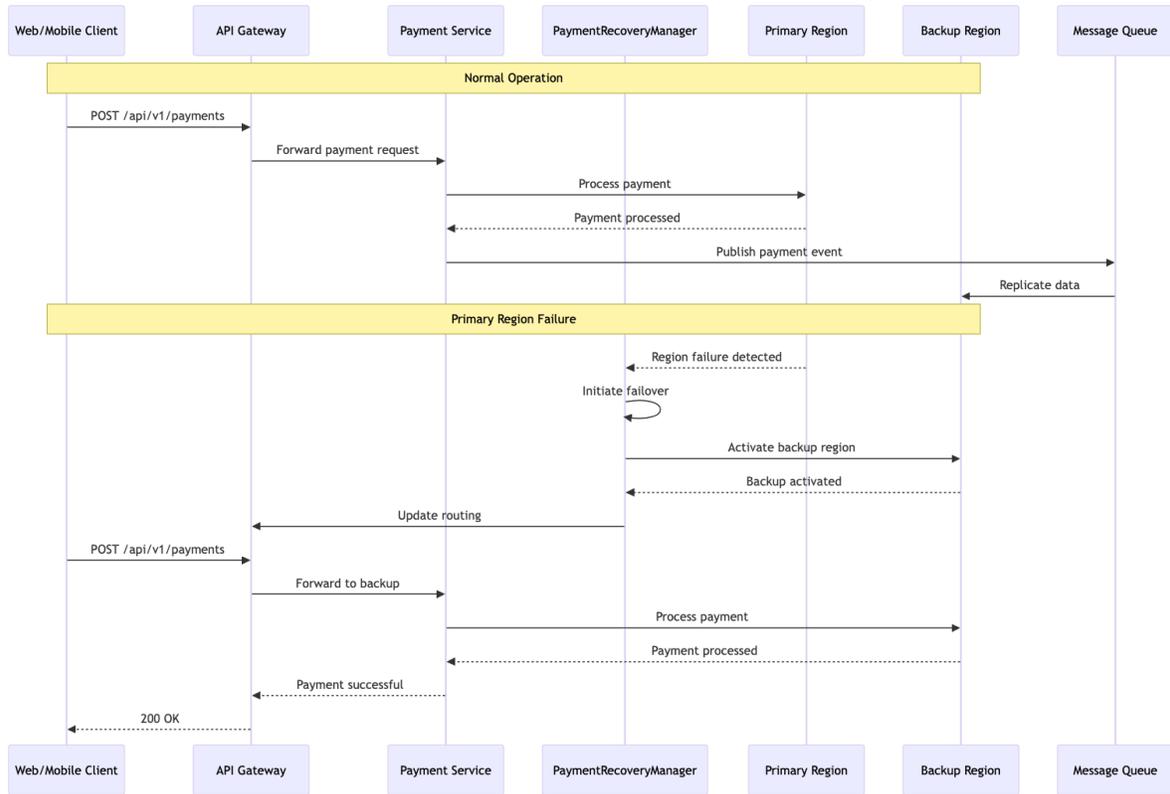

*Figure 12 Sequence diagram for QAS015*

In the case of QAS015 (see Figure 12) a PaymentRecoveryManager initiates a failover, activates a backup region and also requests routes to be updated. This component does not appear elsewhere in the document, but there are other similar components such as TicketRecoveryManager or DeliveryRecoveryManager and a design decision is documented: "Implement RecoveryManager components with failover and backup strategies". This is an example of a discrepancy in the design. Also, the proposed solution was not accepted because it seems like a component within a service is responsible for managing recovery, while this should be a more global approach which is not the responsibility of each service. Although the evaluators observed issues in the design, they mentioned they were impressed by the architecture document and thought this approach could be very useful to practitioners.

We can now provide an answer to RQ2: In our experiment, the LLM produced an architecture design that partially satisfied the drivers. However, in this experiment, the architect did not request the LLM to perform many corrections on the design that is produced. We believed that a careful review and correction of the outputs as they are produced would have resulted in a design that better satisfied the architectural drivers.

## 5.3 Design without ADD

To answer the third research question RQ3 "Does the use of ADD make a significant difference when asking an LLM to design an architecture?", we asked the LLM to perform design without providing the ADD process and associated iteration plan. To explore the extent to which ADD influenced the design outputs we ran three experiments, summarized in table 5. The first experiment employed a zero-shot approach; in the second experiment, we provided an empty architecture document template to the LLM; in the third experiment, we provided this architecture document template and also provided additional instructions in the template such as: "`For each use case and quality attribute scenario, include a sequence diagram that shows how the` components collaborate to achieve the requirement. Below each sequence diagram, add a
short explanation of what is shown in the diagram." Links to the results of the designs are provided
in the result column of the table.

| Approach | Prompt | Result |
|---|---|---|
| Zero-shot | `Consider the requirements described in @ArchitecturalDrivers.md. Design an architecture for this system that satisfies all the requirements described and document this architecture in a document called Architecture.md in the @Design folder. If you include diagrams, use mermaid syntax.` | https://github.com/otrebmuh/HotelPricingSystem/blob/main/OtherDesigns/ArchitectureZeroShot.md |
| Template with the headers of the architecture document | `Consider the requirements described in @ArchitecturalDrivers.md. Design an architecture for this system that satisfies all the requirements described and document this architecture. Document the design in the @Architecture.md document. If you include diagrams, use mermaid syntax.` | https://github.com/otrebmuh/HotelPricingSystem/blob/main/OtherDesigns/ArchitectureEmptyTemplate.md |
| Template with short instructions and empty table. | `Consider the requirements described in @ArchitecturalDrivers.md. Design an architecture for this system that satisfies all the requirements described and document this architecture. Document the design in the @Architecture.md document. Carefully read the lines that start with "Instructions:" in the document and follow these instructions. At the end, remove these lines with instructions. If you include diagrams, use mermaid syntax.` | https://github.com/otrebmuh/HotelPricingSystem/blob/main/OtherDesigns/ArchitectureTemplateWithInstructions.md |

*Table 5. Experiments without providing ADD to the LLM*

The three resulting documents show a design that follows a similar approach of decomposing the system into
microservices for the most important entities, where updates are communicated to the entities using a messaging
system. Some observations about the solutions are the following:
- Solution 1 (zero shot): A Price Publisher Service sends prices to external systems, but this is not correct as the Property Management System and the Commercial Analysis system must read them through an API. Redis was mentioned but it is not shown in diagrams. This resulted in the least detailed design.
- Solution 2 (empty template): This solution contains a similar mistake to the previous solution.
- Solution 3 (Template with instructions): Integration with external systems is not very clear as there is an "Integration Service" that communicates using REST/Events with the external systems. The external systems should go through the API gateway to access this integration service, but it is accessed directly in the proposed solution.

The most comprehensive document is the third one, which shows that adding some instructions to the template and
empty tables makes an important difference. It seems, though, that the LLM did not follow the instructions so
closely as it only included sequence diagrams for the two main use cases.

We can now provide an answer to RQ3: It is clear that using ADD in conjunction with the iteration plan and the
architecture document template makes a significant difference in the resulting design. The design produced using
ADD is much more detailed and also this process gives the opportunity to the architect to make corrections during
the execution of ADD.

# 6. Threats to validity

One of the most important threats to validity in this study is the possibility of the LLM to have "read" the case study from the [Cervantes2024] book during training. We asked the LLM the following prompt: "Are you familiar with the Hotel Pricing System case study from the book "Designing Software Architectures: A Practical Approach" (2nd edition) by Humberto Cervantes and Rick Kazman. If so, do you have details about this case study?" and the response was "I don't have specific details about the Hotel Pricing System case study from the book "Designing Software Architectures: A Practical Approach" (2nd edition) by Humberto Cervantes and Rick Kazman [...] If you're looking for specific details about this case study, you would need to consult the book directly." While we cannot be sure the LLM has not been trained to disclose copyright materials that may have been used during training, it seems to be that the case study was not used during the training process. Furthermore, the proposed solution to the hotel pricing system shows a different approach to what is proposed in the book.

Another threat to validity is that the Event Ticketing System is a type of system which lends itself well as an example. The LLM may have also "read" case studies similar to this one during its training. We indeed found some online examples:

- "Design a Ticket Booking Site Like Ticketmaster"[5].
- "A case study on online ticket booking system"[6]
- "Online events ticketing management system, a case study of Namboole stadium"[7]
- "Scalable Ticketing Platform Development for Good Ticket Company - U.S. Entertainment and Ticketing Brand"[8]

Of these examples, only the Ticketmaster case study goes into deeper technical detail about the architecture. This solution addresses a very similar challenge but many design decisions differ, as shown in Appendix B. These differences allow us to conclude that the solution proposed by the LLM is significantly different. A final threat to validity for this study is that the evaluation with professional architects was made with a very small sample (2 people) from the same company. Also the scenario-based evaluation meeting was constrained in time, so the depth of analysis was limited.

# 7. Analysis and discussion

The two case studies that were used for validation show encouraging results. The architecture documents that were produced in both cases are of high quality and more comprehensive than many examples that we have observed in real projects in industry, and they were produced in a very short time (performing the actual design activity only required a couple of hours each). One key contribution of our approach is the idea of providing an explicit design process to guide the LLM in the design activity. Another important idea is that of asking the LLM to plan design work before performing the actual design through the iteration plan. This is very useful both for the LLM and for the user to know when to stop designing. Without this plan, there is a risk of the LLM producing an excessively complex Big Design Up Front (BDUF). Another fundamental aspect of our approach is the use of a unique architecture document that is modified throughout the design process. In our initial experiments, we documented design using only the iteration documents (following an approach similar to the Hotel Pricing System case study in the book), but this quickly led to problems as each iteration was resulting in structures focused on the drivers of the iteration but there was not an overarching design resulting from these iterations. One limitation of this approach, however, is that as the architecture document grows large, the LLM starts to struggle when making changes to it and this sometimes results in mistakes.

---

[5] https://www.hellointerview.com/learn/system-design/problem-breakdowns/ticketmaster

[6] https://advance.sagepub.com/users/454994/articles/1212909/master/file/data/A%20CASE%20STUDY%20ON%20ONLINE%20TICKET%20BOOKING%20SYSTEM%20PROJECT/A%20CASE%20STUDY%20ON%20ONLINE%20TICKET%20BOOKING%20SYSTEM%20PROJECT.pdf

[7] http://www.researchgate.net/publication/375962782_ONLINE_EVENTS_TICKETING_MANAGEMENT_SYSTEM_A_CASE_STUDY_OF_NAMBOOLE_STADIUM

[8] https://pagepro.co/case-studies/good-ticket-company

One important caveat to consider is that the two solutions used for validation were created using an approach where the user did not take a considerable time reviewing and correcting the LLM about its decisions at the end of each ADD step. This was made on purpose because we wanted to study how good the design could be performed "out of the box". However, there were a number of problems that occurred during the design process and sometimes light corrections were suggested to the LLM. Examples of design problems encountered in the design of the event ticket system are the following.

- When creating the domain model, it did not consider the cloud provider identity service. This resulted in a User entity more complex than needed. The LLM was reminded of the use of the identity service and corrections were made.
- The venue management system, which appears in the context diagram, is not present in the container diagram and is largely ignored in the solution.
- After decomposing several components using a layered approach, the LLM decided to use a hexagonal architecture in the 6th iteration. We mentioned this problem and the LLM accepted the mistake and proceeded to correct it.
- The LLM wanted to introduce a service mesh later in the design. When questioned about the complexity of this approach, it accepted that the same outcome could be obtained by the use of an enhanced api gateway and it proceeded to change the design.

In addition to these design problems, we also faced constant issues related to the LLM "forgetting" information, for example
- It did not present the information as described in step 4 of Attribute Driven Design (see figure 3).
- Frequently forgot to stop after each step and wait for review.
- Forgot to modify the architecture file and started to work on the iteration files exclusively.
- Forgot to adjust the container diagrams.
- Forgot to add design decisions to the architecture document.
- Added sequence diagrams for drivers that were not being addressed as part of the iteration.

Our belief is that many of these problems are related to the amount of context that the LLM receives as part of each prompt. This context is provided by the IDE tool and is beyond our control. In the Event Ticketing System, we also decided to have each user story documented in a separate file and this posed challenges for passing it to the LLM as part of the context. Even though the IDE allows an entire directory to be passed as part of the context, the design decisions show that the user story documents may not have been read by the LLM as expected.

Considering the results we have obtained so far, we believe that it is not yet possible to perform a "*vibe* architecting" approach where we only provide an LLM with requirement information and ask it to produce the design. Architectural design is a complex activity, and performing it automatically is still outside the possibilities of current LLMs (fortunately!). However, the approach we present can be very beneficial if architecture design is performed as a collaboration between the architect and the LLM. The design process can be accelerated significantly, and excellent documentation is produced as part of the design process. One of the difficulties of this approach, however, is that the user that designs in collaboration with the LLM must have enough experience to be able to review and question the decisions made by the LLM. For less experienced architects, there is a risk of blindly accepting the solution "out of the box" and this may result in the introduction of risks or technical debt to the system. Also, LLMs are still prone to hallucination, so reviewing the work they are generating is still a necessary task.

# 8. Conclusions and future work

In this paper we have presented an LLM-assisted approach to assist architects in designing software architectures using the Attribute Driven Design process. Our proposal introduces key aspects such as the use of an iteration plan, an explicit architecture design process (ADD), an architect persona and a single architecture design document which is refined over several ADD iterations. The results we have obtained so far are very encouraging, especially considering that there are few available examples of documented architecture designs. Examples such as the ones in the book [Cervantes2024] were, it appears, not used during the training process as they are copyrighted material. This means that the LLM was capable of making reasonable design decisions and creating and documenting the structures that compose an architecture.

The results in this study give us confidence that it is indeed possible to perform LLM-assisted architecture design, but the process has to be done carefully and in close collaboration between the human and the LLM. It is necessary to review the products of each step of the design process and to correct and guide the LLM. If we omit this crucial step of review and correction of the outcomes of the design process we will get designs that are created with inconsistencies or technical debt from the start.

This paper has presented an initial approach to performing LLM-assisted architecture, and there are many future areas of research:
- **Improving context:** We need to study approaches that improve the information that is passed to the LLM during the design process to avoid omissions and memory losses. This may be challenging when using an IDE as we may not have exact control over what is passed to the LLM and outside the IDE, it may also be challenging as there are associated costs to passing input tokens so there is an incentive to limit context as much as possible to reduce costs.
- **Providing additional design knowledge to the LLM:** Even though LLMs are trained on vast amounts of information, they may not have had access to specialized literature related to designing software architectures, such as the Tactics catalog in [Bass2021]. Providing additional design knowledge may improve the outcomes of the design process. IDEs such as Cursor allow additional documentation to be used by the LLM but it has to be attached as part of a specific prompt. An interesting idea would be to use a specialized architect LLM which is already fine tuned with considerable amounts of architectural and design knowledge.
- **Improve design decisions by making tradeoff analysis:** Evaluating tradeoffs when making design decisions is essential. Currently, we are not explicitly asking the LLM to analyze tradeoffs when making its design decisions, and studying how well the LLM can make this type of analysis remains to be done.
- **Evaluating architectures using LLMs:** An interesting idea is to use a different LLM to evaluate the output of the design process. The architecture document facilitates this task enormously and we have already made some informal experiments with encouraging results, but a more formal approach still needs to be investigated. A sample analysis is provided in the the Analysis folder of both projects presented in this paper[9] and [10]. An interesting possibility is to perform an automated evaluation at the end of every iteration and provide the feedback from the evaluating LLM to the designing LLM.
- **Generating code from the design:** The next logical step from the design process is to generate the actual system code by giving the LLM the architecture document. Similar to the evaluation activity, we have started experimenting this approach with the HPS design. We have observed discrepancies between code and design (probably due again to a problem with prompt context) so we still need to study how to improve this approach with more detail.
- **Continuous changes to the design:** The design approach presented here is not a Big Design Up Front approach. Ideally, a few ADD iterations are performed at the beginning of a project in a design round and additional iterations are performed later in other iteration rounds, as the system is being built. We need to explore the process of performing continuous design iterations rounds and how these impact the documentation and the code.
- **Updating the architecture document from changes in the code:** Ideally the architecture document should always reflect the state of the code, however, it may happen that changes are done to the code without the architecture documentation being updated. This can be problematic if we intend to perform additional ADD iterations on an outdated architectural model. Automatically updating the architectural document from changes in the code is an interesting research area.
- **Use of other LLMs:** the experiments conducted in this paper were done using the same LLM: Claude Sonnet 3.7. It may be that other LLMs may produce better results so a comparison remains to be done.

Finally, it is important to consider whether or not it is desirable to reach a point where LLMs can design software architectures automatically. Our opinion, at this time, is that design is a creative act that is better performed as a collaboration between a human and an LLM.

---

[9] https://github.com/otrebmuh/HotelPricingSystem/blob/main/Analysis/AnalysisResults.md
[10] https://github.com/otrebmuh/EventTicketSystem/blob/master/Analysis/AnalysisResults.md

# Appendix A

The following table presents a comparison between the solution produced by the LLM and the one that is documented in the case study of [Cervantes2024].

| Aspect | Case study | LLM generated solution |
| --- | --- | --- |
| **System Structure** | The design splits the system into a Command service (for writes) and Query service (for reads) following Command Query Responsibility Segregation (CQRS), plus an Export service for integration. These services are deployed behind a single API Gateway in a private cloud network. The Command and Query components communicate via an event bus (Kafka) instead of direct calls. This results in only a few core back-end services (command, query, export) that handle all pricing functionality. | The design uses a microservices approach but with more fine-grained separation. In addition to Price Management (command) and Price Query (read) services, there are distinct services for Hotel Management, Rate Management, and Authentication. An API Gateway fronts all these services. An Event Bus enables asynchronous communication between services. The existence of the Query service is essentially an implementation of CQRS. |
| **Components** | Each microservice in the design is structured using standard Spring layered architecture. This mapping of entities to controllers/services ensures all required modules are identified for each use case. The focus is on fundamental components needed for functionality, with security and basic logging considered. | Each service in the design contains multiple specialized components additional to the standard Spring Layers components (e.g. Calculation Engine, Event Store, Event Publisher with Outbox, Circuit Breaker). This design is more complex and incorporates patterns for reliability and performance. |
| **Sequence diagrams** | The case study presents a limited number of sequence diagrams to illustrate certain functional and non functional scenarios | The architecture document includes sequence diagrams for all functional and quality attribute scenarios. |
| **Contracts** | In iteration 2, a few container and component-level contracts are presented, mostly for illustration purposes. | The architecture document includes detailed container-level REST contracts for the different services. Endpoints are described in tables with methods, descriptions, request body and response.<br><br>Additionally, description of events and event payloads is also included. |
| **Communication Style** | The design emphasizes asynchronous communication. The Command service publishes price change events to an event bus, which the Query and Export services subscribe to. External interactions use synchronous REST calls via the API Gateway, but internal services exchange data through events. | The design employs an event-driven architecture for inter-service communication. Services publish events (hotel updates, rate changes, price changes) to a centralized Event Bus and subscribe where needed. The architecture also provides synchronous APIs through the API Gateway for clients. The Query service supports multiple protocols: both REST and gRPC APIs are offered for retrieving prices. |
| **Domain Modeling** | In iteration 2, an explicit domain model is defined for the Hotel Pricing context. Key domain entities include Hotel, RoomType, Rate, and Price, along with business rules for price calculations. This domain model is confined to the command side (write model). The Query service does not maintain an object model of its own; it simply stores price events in a database for fast retrieval, and the Export service likewise has no domain data (just passes events along). Thus, the case study design treats "hotel pricing" as one domain context, centrally modeled in the command microservice. | The design models similar core concepts (Hotel, Rate, RoomType, Price, BusinessRule, etc.), but splits responsibility across services. This design follows a Domain-Driven Design approach by dividing the domain model across context-specific services. |
| **API Management** | The case study employs an API Gateway as the single entry point for all clients (users or external systems). The Gateway forwards requests to the appropriate back-end microservice. Each microservice exposes RESTful endpoints for its functionality. Authentication and basic authorization are enforced at the gateway and service level (integrating with the cloud identity service). There is no mention of alternate protocols or versioning strategies in the early design; the focus is on providing the necessary REST endpoints for each use case. API security (e.g. token validation) and request validation tactics are considered as part of securing these APIs. | The design also uses a central API Gateway that routes requests to the appropriate service and applies cross-cutting concerns like auth, rate limiting, and input validation. Each microservice defines a set of APIs (documented in the architecture's Interface section) for its domain. The design explicitly supports API versioning and multiple protocols. In particular, the Price Query service offers both a REST API and a gRPC interface for clients with different needs. |
| **Security** | User login (HPS-1) is handled by leveraging the company's cloud-based User Identity Service for Single Sign-On. When a user provides credentials, HPS delegates verification to this identity service and, upon success, grants access with appropriate permissions. Each API call must be authenticated (e.g., via tokens issued by that identity service). Authorization checks are mentioned as a requirement, but how this is handled is not discussed. In iteration 1, security tactics such as authenticating and authorizing actors, limiting access to resources, and encrypting API data were identified as critical and were planned from the start. | The design introduces an Authentication Service as a dedicated microservice to handle auth concerns. The API Gateway delegates authentication requests to this Auth service. The Auth service in turn integrates with the same cloud identity provider via OAuth2/OIDC protocols to validate credentials and tokens. It likely manages user sessions or JWT tokens and also stores authorization data. The design explicitly mentions using the cloud identity (CON-2) for user management and SSO, similar to the case study approach, but here it's encapsulated in a service. |
| **Scalability** | A primary motivation for the case study's CQRS and microservice design was to address scalability and performance. By splitting the query side from the command side, the system can scale reads independently of writes. In iteration 3, the design explicitly adds replication for the Query service and load balancing to increase throughput for price queries. Kafka as an event store also allows scaling consumers (multiple query or export processors) if needed. This approach ensures that heavy read load from external systems querying prices can be met by adding resources to the Query microservice alone. Other scalability considerations include avoiding shared state (each microservice has its own database) and using stateless services to facilitate horizontal scaling. | The architecture considers independent scaling of each microservice. For instance, if price queries spike, the Price Query Service can be scaled to many instances without touching the others. The inclusion of a distributed caching layer in the Query service supports high read throughput by reducing database load. Additionally, the architecture accounts for scaling in multiple dimensions: it uses CQRS to divide read/write load, and it can leverage multi-region deployment to distribute load and reduce latency for users in those regions. The decision to containerize everything and orchestrate via Kubernetes means the platform can auto-scale services based on demand. |
| **Deployability** | The case study considers deploying all components on a cloud provider. In iteration 1, it was decided to keep all microservices and the event bus within a private network, accessible only through the API Gateway (enhancing security). By iteration 3, more concrete deployment tactics are defined: using a container orchestration service (likely a managed Kubernetes) to run and monitor the microservices, and setting up passive redundancy (a standby deployment in a second availability zone/region) to take over on failures. The event bus (Kafka) is a managed, durable service, removing the need to manage message brokers manually. Each microservice has its own database (e.g., a SQL DB for Command, NoSQL for Query) provided as managed cloud databases. Continuous deployment (CI/CD) is satisfied using a proprietary git-based platform with support for pipelines. | The architecture specifies a robust deployment and DevOps strategy. All services are packaged as containers and deployed on a container orchestration platform (Kubernetes) for consistency across environments. A continuous integration/continuous deployment pipeline builds images, runs tests, and promotes deployments through staging to production. Configuration is managed separately (Infrastructure as Code and a config repository) to enable reproducible, environment-specific deployments. For high availability, the system is deployed in multiple regions with automated failover: if the primary region fails, a secondary region's services and database replicas take over, using global DNS or load balancer routing. The design also leverages managed cloud services (for databases, identity, etc.) to offload infrastructure management. |
| **Testability** | The use of the Spring framework and its inversion of control was selected to facilitate unit testing of modules. A concern (CRN-7) was added to ensure all custom business logic components can be unit tested. This implies the design will include a suite of unit tests for services, controllers, etc., and likely some integration tests for the event flow. However, the case study description does not describe a full testing environment setup. No specific mention is made of test automation tools or frameworks beyond standard unit testing. | The design dedicates an entire infrastructure for testing. It uses containerized test environments via TestContainers to spin up ephemeral databases, Kafka brokers, and other dependencies for integration tests. There is a suite for contract testing to ensure that each service's APIs meet the expectations of its consumers (using consumer-driven contracts and schema validation). Mock versions of external systems are provided to test interactions without relying on live systems. The test infrastructure supports automated provisioning of these components to run integration tests in CI pipelines. Additionally, test data management tools reset and seed databases between tests to ensure isolation. All of this is integrated into the CI/CD pipeline, so tests run on each build and deployment, with reporting and failure alerts. |

*Table A. Design comparison between published case study and LLM solution*

# Appendix B

The following table compares the solution generated by the LLM with the ones from the online solution.

| Area | Ticketmaster Article | LLM generated solution |
| --- | --- | --- |
| **Service Granularity** | Small set of coarse-grained services (API Gateway, Event, Search, Booking, Payment) | Fine-grained microservices: Auth, User, Event, Search, Inventory, Order, Payment, Ticket, Delivery, Notification, etc. |
| **Seat Reservation Strategy** | Prefers a distributed lock with TTL in Redis to prevent double-booking | Uses optimistic locking in the Inventory DB plus cache invalidation and event sourcing |
| **Peak-Demand Mitigation** | Suggests Server-Sent Events for live seat updates and an optional virtual waiting room built on WebSockets | No explicit real-time push or waiting-room mechanism; relies on caching and eventual consistency via MQ events |
| **Search Synchronisation** | Change Data Capture (CDC) pipeline from PostgreSQL to Elasticsearch to keep the index fresh in near real time | Event Service publishes **"EventCreated"** messages; Search Service consumes and indexes, Event-driven rather than CDC |
| **Ticket Generation & Delivery** | Out of scope—design stops once booking is confirmed. | Dedicated Ticket Service and Delivery Service with multi-channel strategy (email / mobile) |
| **Security & Compliance** | Assumes Stripe handles payment security; little depth on broader security. | Explicit PaymentSecurityManager, encryption, circuit-breaker around gateway, JWT/OAuth via external IdP |
| **Data Stores** | Single relational DB (PostgreSQL) plus Redis & Elasticsearch. | Dedicated DB per service, read-replicas, potential sharding and audit logs per domain |

*Table B*. Solution comparison